\documentclass[conference]{IEEEtran}
\IEEEoverridecommandlockouts
\usepackage{cite}
\usepackage{amsmath,amssymb,amsfonts}
\usepackage{algorithmic}
\usepackage{graphicx}
\usepackage{textcomp}
\usepackage{xcolor}
\def\BibTeX{{\rm B\kern-.05em{\sc i\kern-.025em b}\kern-.08em
    T\kern-.1667em\lower.7ex\hbox{E}\kern-.125emX}}
\begin{document}

\title{Real-Time Machine Learning Enabled Low-Cost Magnetometer System\\

\thanks{This work was supported by NSF EPSCoR Award OIA-1920965.}
}

\author{\IEEEauthorblockN{Talha Siddique and Md. Shaad Mahmud*}
\IEEEauthorblockA{\textit{Department of Electrical and Computer Engineering} \\
\textit{University of New Hampshire}\\
Durham, NH, USA \\
*mdshaad.mahmud@unh.edu}
}

\maketitle

\begin{abstract}
Geomagnetically Induced Currents (GICs) are one of the most hazardous effects of space weather. The rate of change in ground horizontal magnetic component $dB_{H}/dt$ is used as a proxy measure for GIC. In order to monitor and predict $dB_{H}/dt$, ground-based fluxgate magnetometers are used. However, baseline correction is crucial before such magnetometer data can be utilized. In this paper, a low-cost Machine Learning (ML) enabled magnetometer system has been implemented to perform real-time baseline correction of magnetometer data. The predicted geomagnetic components are then used to derive a forecast for $dB_{H}/dt$. Two different ML models were deployed, and their real-time and offline prediction accuracy were examined. The localized peaks of the predicted $dB_{H}/dt$ are further validated using binary event analysis.
\end{abstract}


\section{Introduction}
Geomagnetically Induced Currents (GICs) are current induced on the Earth's surface conductors due to Geomagnetic Disturbances (GMDs) or Geostorms \cite{camporeale_machine_2018} \cite{pirjola_geomagnetically_2000}. GMDs occur due to the interaction of charged particles from the Sun with the Earth's Magnetosphere \cite{lakhina_geomagnetic_2016}. The flow of GIC into power transmission lines can lead to the malfunction of electrical power devices, and transformers \cite{rajput_insight_2021}. Therefore, to get an adequate opportunity to mitigate the risks posed by future Geostorms, there is a need to analyze and predict GIC in near real-time. \cite{morley_perturbed_2018} \cite{LUNDSTEDT20052516}.

GICs are difficult to estimate, and the rate of change in the local ground horizontal magnetic component ($dB_{H}/dt$) is used as a proxy value \cite{Keesee2020}. The disturbances in the geomagnetic field are monitored through the use of ground-based magnetometers. However, such devices' ground magnetic component readings tend to suffer from background noise and interference \cite{Keesee2020}. For such data to be utilized for GIC prediction, initiatives have been undertaken where magnetometer data are curated after baseline correction. For example, SuperMAG is an association of organizations that operates more than 300 ground-based magnetometers. Their objective is to perform baseline removal from the collected data and disperse them for the wider scientific community \cite{2012JGRA..117.9213G}. Also, there is no scientific consensus regarding a common technique for baseline correction \cite{2012JGRA..117.9213G}. The task mentioned above usually includes performing complex mathematical operations that are offline in nature. Because of this, the dispersion of the processed data is time-consuming. The literature consists of studies where machine learning (ML) architectures like Convolutional Neural Network (CNN) have been implemented for baseline removal \cite{schmidt2019peak}. However, these techniques do not process the data in real-time. With the availability of raw data, there is also a drive within the space science community to develop Artificial Intelligence (AI)-enabled hardware \cite{nasa2021report}. In recent years, advancements have been made in the Internet of Things (IoT) and Single Board Computer (SBC) technology, where the data is partly processed on edge devices. However, such technology is still in its infancy \cite{yousuf_ibug:_2022}. To address these gaps, a low-cost ML-enabled magnetometer system has been developed that can predict baseline-corrected magnetic field component values in real-time. Section-II in this paper covers the methodology, which describes the hardware setup, data acquisition and processing, and an overview of the implemented model. Section-III provides an analysis of the obtained results. Finally, section-IV concludes the paper and discusses the future possibilities.

\section{Methodology}

\subsection{Hardware and Experimental Setup}

\begin{figure}
\begin{center}
\includegraphics[width=7cm,height=3.5cm]{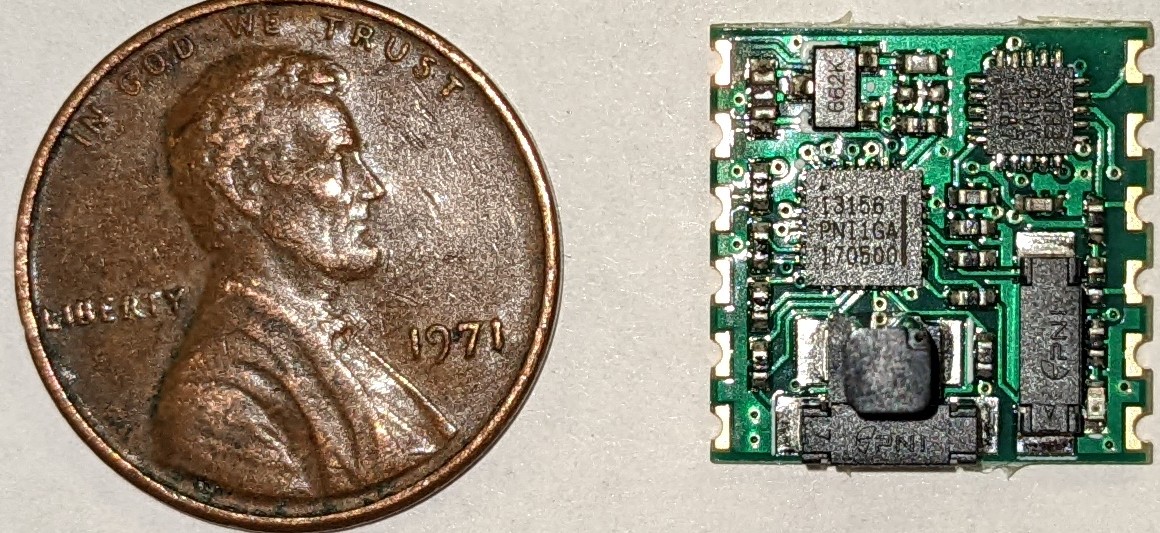}
\end{center}
\caption{PNI RM3100 magnetometer sensor shown next to a US cent coin for size comparison.}\label{fig:1}
\end{figure}

\begin{figure}
\begin{center}
\includegraphics[width=7cm,height=4cm]{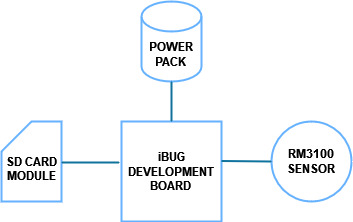}
\end{center}
\caption{Schematic diagram of hardware setup, with key hardware components labelled}\label{fig:2}
\end{figure}

For this study, an iBUG development board was connected to an RM3100 magnetic sensor and an SD card module to store the raw and predicted magnetic component data. These devices were connected to the development board using the Serial Peripheral Interface (SPI) protocol. The board was powered via its USB port by an external portable power pack. The power pack had a capacity of 10000 mAh and an output of 5V. The iBUG board is an ML enabled IoT sensing platform, which has been used in the past literature for real-time environmental monitoring \cite{yousuf_ibug:_2022}. It has a RAK11300 module with a 133MHz processor and a dual-core Raspberry RP2040 Microcontroller Unit (MCU). RM3100 is the propriety magneto-inductive geomagnetic sensor developed by Positioning Navigation Intelligence (PNI) Corporation \cite{Leuzinger2010}. Magneto-inductive based sensors have been deemed a promising technology to study space and ground-based geomagnetic activity \cite{regoli_investigation_2017}. Figure-\ref{fig:1} depicts a picture of the RM3100 sensor. In addition, the sensor was chosen due to its low cost and compatibility with development boards like iBUG. The entire hardware setup costs around \$90, whereas the average cost of a professional fluxgate magnetometer system is about \$1100 \cite{2020AGUFMSM011..12S}. The schematic diagram of the experimental setup is illustrated in Figure-\ref{fig:2}.

\subsection{Data Acquisition and Processing}

This paper used the baseline-corrected $X'$ and $Y'$ component data from a fluxgate magnetometer as the model's target variables. The raw data from the RM3100 sensor, $X$ and $Y$, were used as input variables. University of New Hampshire (UNH) Space Weather Underground (SWUG) is a collaboration between the UNH and Northern New England high schools to assemble and deploy "3-Axis Simple Aurora Monitor (SAM-III)" fluxgate magnetometers, hereinafter referred to as SWUG magnetometer. This initiative aims to collect geomagnetic data to study the impact of GMDs, and GICs \cite{2020AGUFMSM011..12S}. The data collected by SWUG on a given day is processed for missing data and released on its website 24 hours later, along with its unprocessed counterpart \cite{gssc}. The SWUG deployed magnetometer nearest to our location was in Exeter, NH. The experimental setup described in the previous sub-section was placed in Exeter, NH but not in the same geo-location as the SWUG magnetometer.

The input data was collected using the RM3100 sensor at 2 seconds interval for 16 hours. The correlation between the $X$ and $Y$ component data of both the SWUG magnetometer and the RM3100 sensors was examined. This was done to check the calibration difference between the SWUG magnetometer and the RM3100 sensor. The $X$-components have a correlation of 0.53, whereas the $Y$-component has a correlation of 0.79. Because of this difference in calibration, the raw data from the RM3100 were used as the input variable instead of the data from the SWUG magnetometer. The data for the output variables $X'$ and $Y'$ were derived from the SWUG magnetometer component data by performing a baseline-correction. As mentioned in Section-I, there is no consensus within the scientific community regarding a common methodology for baseline correction. Therefore, this paper used Asymmetric Least Square (ALS) for baseline fitting. The ALS algorithm aims to match the baseline with the measured data while simultaneously penalizing the roughness of the baseline \cite{peng2010asymmetric}. A general objective function of ALS is exhibited in Equation-\ref{eq:1}, where $a_{i}$ represents the data; $z_{i}$ is the estimated baseline; $w_{i}$ is the weight; $\lambda$ is the penalty factor; and $\Delta^d$ represents the finite difference operator of order $d$. A python library called "pybaselines" was used to carry out the baseline-correction on the $X$ and $Y$ components from the SWUG magnetometer data.

\begin{equation}\label{eq:1}
    \sum\limits_{i= 1}^N w_{i}(a_{i} - z_{i})^2 + \lambda\sum\limits_{i= 1}^{N-d} (\Delta^d z_{i})^2
\end{equation}

The input and output variable datasets were void of missing data points. The dataset was split for training and testing, where 80\% of the data was used for offline training, and the remainder was used for offline testing. The validation set was created by taking 20\% of the data from the training dataset.

\subsection{Model and Deployment Overview}

Two different, off-the-shelf Deep Learning (DL) architectures were implemented- a feed-forward Artificial Neural Network (ANN) and a Convolutional Neural Network (CNN). Both the Neural Networks (NN) has three hidden layers and use the Adam optimizer to minimize Root Mean Square Error (RMSE) as the loss function. Both the models are trained separately offline, using the training and validation set. The models are multivariable and multivariate, as they take $X$ and $Y$ components from RM3100 as input and predict the baseline-corrected $\hat{X}$ and $\hat{Y}$ as output. The implemented magnetometer system then derives the forecasted rate of change in the local ground magnetic horizontal component (${d\Hat{B}_{H}/dt}$). The derivation of ${d\Hat{B}_{H}/dt}$ is exhibited in Equation-\ref{eq:2}, where $dt$ is over a time unit of 1 second \cite{Keesee2020}. Tensorflow Lite (TF-Lite) version of the models was generated, and the TF-Lite models were deployed at different instances to test their real-time prediction capability. The experimental setup was placed in the same geo-location as before to test the real-time prediction capability. For each deployed model, 2 hours of real-time predicted data were collected.

\begin{equation}\label{eq:2}
\frac{d\Hat{B}_{H}}{dt} = \sqrt{(\frac{d\Hat{X}}{dt})^2 + (\frac{d\Hat{Y}}{dt})^2}
\end{equation}

\section{Results and Discussion}

The trained model was tested offline using the test dataset. Table-\ref{table:1} summarizes the accuracy of the real-time and offline prediction of the two components for each model. Normalized RMSE was used as the error metric. It can be observed that in the case of real-time prediction, the CNN model outperforms the ANN model. However, the two models have the same accuracy for offline prediction. Also, the offline model performs better relative to its real-time counterpart. The offline model performs better because the training and test set was collected on the same day. Given the lack of variation in the dataset, the trained model did not achieve the required level of generalization to provide a comparable real-time prediction carried out on a different day.

\begin{table}[htbp]
\caption{Real-time and offline model accuracy}\label{table:1}
\vspace{0.008cm} 
\begin{center}
\begin{tabular}{|c|c|c|}
\hline
\cline{2-3} 
\textbf{\textit{MODEL}} & \textbf{\textit{REAL-TIME RMSE}}& \textbf{\textit{OFFLINE RMSE}}\\
\hline
ANN ($X$) & 0.60 & 0.29  \\
ANN ($Y$) & 0.61 & 0.30 \\
CNN ($X$) & 0.67 & 0.28 \\
CNN ($Y$) & 0.68 & 0.28 \\
\hline
\end{tabular}
\label{tab1}
\end{center}
\end{table}
\vspace{0.01cm}

\begin{figure}
\includegraphics[width=8.5cm,height=7cm]{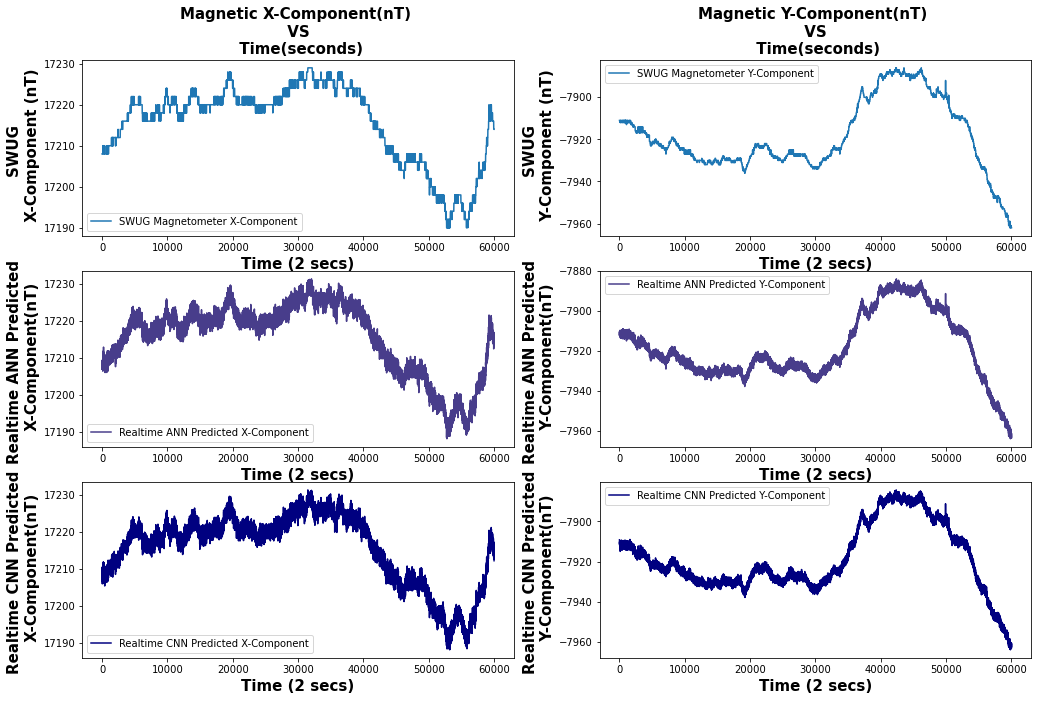}
\caption{Magnetic component ($X$ and $Y$) versus time}\label{fig:3}
\end{figure}
\vspace{0.01cm}

Figure-\ref{fig:3} illustrates the offline baseline-corrected SWUG magnetic component data ($X'$ and $Y'$ ) and their real-time predicted counterparts ($\hat{X}$ and $\hat{Y}$ ) over time. For a given time frame, the $\hat{X}$ and $\hat{Y}$ predicted by the ANN and CNN have a greater degree of fluctuation than $X'$ and $Y'$. The high RMSE values for the real-time prediction stem from these fluctuations. Despite the high RMSE, both the ANN and CNN predictions were able to capture the trend of the baseline-corrected data collected from the SWUG magnetometer.

The past literature on space science recommends using the Heidke Skill Score (HSS) to measure a model's performance \cite{pulkkinen_community-wide_2013}. The HSS measures the fractional improvement of the forecast over the standard forecast. It is also a means of validating how well the predicted data captures the time-series trend of the ground truth. HSS is determined through a binary event analysis, where four thresholds (0.3 nT/s, 0.7 nT/s, 1.1 nT/s, 1.5 nT/s) are used as metrics for validating predictions of $dB_{H}/dt$ \cite{Keesee2020} \cite{pulkkinen_community-wide_2013}. The actual and forecasted time series $dB_{H}/dt$ values are divided into non-overlapping, 1200 second windows, and for each window the local maximum $dB_{H}/dt$ is determined. For a given threshold, the actual and forecasted local maximum $dB_{H}/dt$ are examined to determine whether they cross the threshold or not. From the aforementioned analysis, the number of true positives (H), true negatives (N), false positives (F), and false negatives (M) can be determined. The HSS for a particular threshold is then determined using Equation-\ref{eq:3} \cite{pinto_revisiting_2022}. 

\begin{equation}\label{eq:3}
 HSS = \frac{2(HN - MF)}{(H+M)(M+N)+(H+F)(F+N)}
\end{equation}

The HSS for the implemented models is shown in Table-\ref{table:2} for each of the four threshold values. HSS ranges from $-\infty$ to 1, where a negative value indicates that the model forecasts better by chance; 0 means the model has no skill; 1 represents perfect skill. As shown in Table-\ref{table:2}, both the models have positive skill scores, with the ANN model performing marginally better than CNN. Given that the models could capture the general trend with positive skill scores, this implemented system lays the foundation for real-time peak detection of forecasted $dB_{H}/dt$, which is crucial for predicting GMDs.

\begin{table}[htbp]
\caption{Heidke Skill Score(HSS) for $dB_{H}/dt$ forecasted by each model}\label{table:2}
\begin{center}
\begin{tabular}{|c|c|c|}
\hline
\cline{2-3} 
\textbf{\textit{THRESHOLD (nT/s)}} & \textbf{\textit{HSS (ANN $dB_{H}/dt$)}}& \textbf{\textit{HSS (CNN $dB_{H}/dt$)}}\\
\hline
0.3 & 0.47 & 0.41 \\
0.7 & 0.38 & 0.32 \\
1.1 & 0.28 & 0.23 \\
1.5 & 0.18 & 0.10 \\
\hline
\end{tabular}
\label{tab1}
\end{center}
\end{table}

\vspace{0.001cm}
\section{Conclusion and Future Work}

This paper implements a low-cost ML-enabled magnetometer system that can predict real-time baseline correction of $X$ and $Y$ geomagnetic components. The system used an RM3100 magneto-inductive sensor connected to an iBUG development board. Two different ML architectures were deployed, and their real-time and offline prediction accuracy were examined against the baseline removed data from a fluxgate magnetometer. Although the offline model outperforms the real-time model, the real-time model captured the general trend of the ground truth data from the fluxgate magnetometer. This, in turn, opens up the possibility of peak detection in real-time using ML, which is a crucial process for GMD detection. Going forward, distributed real-time ML architectures like federated learning will also be explored. 

\section*{Acknowledgment}

We thank members of the the “Machine-learning Algorithms for Geomagnetically Induced Currents In Alaska and New Hampshire” (MAGICIAN) team. This work was supported by NSF EPSCoR Award OIA-1920965.

\bibliographystyle{IEEEtran}
\bibliography{conference_101719}

\end{document}